# Thin-film 'Thermal Well' Emitters and Absorbers for High-Efficiency Thermophotovoltaics

Jonathan K. Tong[1], Wei-Chun Hsu[1], Yi Huang[1], Svetlana V. Boriskina[1,*], and Gang Chen[1,*]

[1]Department of Mechanical Engineering, Massachusetts Institute of Technology, Cambridge, MA 02139

**Abstract**

A new approach is introduced to significantly improve the performance of thermophotovoltaic (TPV) systems by using low-dimensional thermal emitters and photovoltaic (PV) cells. By reducing the thickness of both the emitter and the PV cell, strong spectral selectivity in both thermal emission and absorption can be achieved by confining photons in trapped waveguide modes inside the thin-films that act as thermal analogs to quantum wells. Simultaneously, photo-excited carriers travel shorter distances across the thin-films reducing bulk recombination losses resulting in a lower saturation current in the PV cell. We predict a TPV efficiency enhancement with near-field coupling between the thermal emitter and the PV cell of up to 38.7% using a germanium (Ge) emitter at 1000 K and a gallium antimonide (GaSb) cell with optimized thicknesses separated by 100 nm. Even in the far-field limit, the efficiency is predicted to reach 31.5%, which is an order of magnitude higher than the Shockley Queisser limit of 1.6% for a bulk GaSb cell and a blackbody emitter at 1000 K. The proposed design approach does not require nanoscale patterning of the emitter and PV cell surfaces, but instead offers a simple low-cost solution to improve the performance of a thermophotovoltaic system.

*Correspondence and requests for materials should be addressed to either S.V.B. or G.C. (emails: sborisk@mit.edu, gchen2@mit.edu)

A thermophotovoltaic system is a unique type of heat engine that directly converts thermal energy to electricity.[1–4] Typical TPV systems consist of a thermal emitter and a photovoltaic cell in which there are no moving parts allowing for a compact power generation platform. In theory, TPV systems can convert radiative energy from the thermal emitter to electricity at an efficiency approaching the Carnot limit for monochromatic radiation and can be used for both waste heat and solar energy harvesting.[5–13] However, the efficiency achieved in practical TPV systems has been limited by the mismatch between the thermal emission spectra and the PV cell absorption, thermalization of high energy charge carriers, and non-radiative recombination losses which are typically high in narrow band gap PV cells used for TPV.

To improve the efficiency, past studies have generally followed two approaches: (1) using intermediate frequency filters, such as rugate filters, photonic crystal filters, and plasma reflectors, which recycle low energy photons by returning them back to the thermal emitter and (2) engineering the emitter using rare-earth dopants, plasmonic metamaterials, and photonic crystals to achieve spectrally selective emission and angular selectivity.[14–24] By utilizing these approaches and minimizing system losses, several recent studies have demonstrated significant improvement to overall system efficiency for combustion and solar-based TPV systems.[25–28] In addition, previous studies have also exhibited TPV efficiencies, defined as the electrical power output divided by the net radiative heat input from the emitter to the PV cell, of nearly 24% by utilizing ternary, quaternary, and multi-quantum well cells at emitter temperatures of 1300-1500 K.[29–32] In parallel, other theoretical studies proposed utilizing the strong near-field coupling of electromagnetic fields between the emitter and PV cell separated by nanometer gaps to improve the efficiency and power density of TPV systems.[33–40] However, the proposed designs for high-efficiency near-field TPV systems required extremely narrow gap separations between the emitter and the PV cell, ultra-low band gap PV cells, and high emitter temperatures limiting practical realization.

In this work, we present a new approach to improve the spectral selectivity of a TPV system that is based on reducing the dimensions of the emitter and the PV cell in order to spectrally shape emission and absorption in analogy to the use of electronic quantum confinement effects in quantum wells.[41,42] In lieu of the electronic comparison, we define this approach henceforth as the 'thermal well' effect. The proposed design consists of a thin-film thermal emitter and a thin-film PV cell, which support quantized waveguide modes trapped inside the material by total internal reflection. The dispersion characteristics of these waveguide modes lead to resonant enhancement in the density of optical states (DOS) resulting in greater thermal emission and absorption especially in the near field.[43,44] More importantly, these waveguide modes also exhibit cut-off frequencies beyond which thermal radiation is suppressed since no mode is available for emission and absorption.[45,46] For an optimally thick thin-film emitter and PV cell, emission and absorption will be resonantly enhanced for photons with energies larger than the band gap of the cell and suppressed for photon energies smaller than the band gap. In this manner, the use of simple morphological structuring effects can dramatically improve the spectral selectivity of a TPV system. In addition, bulk non-radiative recombination losses are also reduced in the thin-film PV cell resulting in a lower saturation current thus improving the electrical performance of the cell. The combination of these effects is predicted to result in a high TPV energy conversion efficiency at relatively low emitter temperatures. This effect can also be observed in both the far-field and near-field regimes where evanescent wave coupling of the trapped optical modes can further increase the conversion efficiency.

## Results

**Shaping Radiative Heat Transfer with Thermal Wells.** The TPV system considered in this study consists of two parallel thin-films, representing the emitter and the PV cell, which are separated by a gap g as shown in Fig. 1a. The emitter and the PV cell are at uniform temperatures of $T_H$ and $T_C$, respectively. The view factor is assumed equal to 1. It is also assumed that the materials of both the emitter and the cell have isotropic and local dielectric permittivities. Both of the thin-films are placed onto semi-infinite metallic back reflectors. In this work, back reflectors made of an ideal perfect metal and a real metal will be considered. The use of a perfect metal eliminates emission and absorption in the substrate thus providing a way to estimate the performance of the proposed TPV design due to only the thermal well effect. These results can be directly compared to the previously studied case of a bulk emitter and a bulk PV cell to highlight the benefits of morphological structuring.

In this study, thermal radiation is modelled using a rigorous analytical electromagnetic formulation based on Rytov theory (see Methods).[47–49] Several configurations are evaluated to assess the potential of using thin-film emitters and absorbers to improve TPV performance. For all cases, a GaSb PV cell with a band gap of 0.726 eV is used, which is conventional in many TPV platforms.[50–53] To demonstrate the thermal well concept, we chose a thin-film of Ge as the thermal emitter. Ge is a high refractive index semiconductor with a band gap of 0.7 eV. The material absorption in Ge at wavelengths above the band gap naturally provides strong emission channels by virtue of Kirchoff's law. By combining the spectral properties of bulk Ge absorption with morphological structuring of the emitter, it is possible to spectrally tailor thermal emission for TPV systems.

We also considered a bulk W emitter in order to compare thermal emission spectra of waveguide modes confined in Ge thermal wells with previously studied emission of surface plasmon polariton (SPP) modes thermally excited on the W surface.[35] W is a refractory metal that is thermally stable at high temperatures and supports SPP modes in the near infrared wavelength range. SPP modes are resonant surface waves that form between materials with dielectric permittivities of opposite sign. These modes exhibit significant enhancement in DOS and hence radiative transfer when in close proximity to the surface.[35,43] However, due to the highly confined nature of SPP modes, this enhancement in DOS decreases rapidly away from the surface. Nevertheless, W emitters are still used in more conventional TPV systems because of its low plasma frequency.[54]

The backside metal supporting the emitter and the PV cell is chosen to be either a perfect metal (PM), silver (Ag) on the PV cell side, or tungsten (W) on the emitter side. When both W and Ag are used, a magnesium fluoride ($MgF_2$) spacer is placed between the GaSb cell and the Ag mirror in order to minimize coupling of SPP modes in W and Ag in the near-field regime. The optical constants of GaSb, Ge, $MgF_2$, Ag and W were obtained from literature.[55] The infrared optical properties of Ag were extrapolated using a Drude model.

To demonstrate the effect of reducing a bulk system to a thin-film structure, Figs. 2a and 2b shows the normalized transmission function, $G_{NT}$, for a bulk Ge emitter and a bulk GaSb cell and their corresponding thin-film counterparts, respectively. For the thin-film thermal well structure, a perfect metal was assumed to be on the backside of both the emitter and the PV cell. The optimum film thicknesses that maximize the TPV efficiency were obtained assuming an emitter temperature of 1000 K, a PV cell temperature of 300 K, and a gap separation of 100 nm. The emitter temperature was chosen to be below the melting temperature of Ge which is nearly 1200 K. As shown in Fig. 2, the transmission function for the bulk system indicates that radiative modes are supported over a broad frequency range. This corresponds to the $n^3$ enhancement in the bulk photon DOS of high refractive index media.[43] In contrast, the transmission function for the thin-film structure changes dramatically and is now comprised of several distinct bands which indicate the presence of trapped waveguide modes in both the Ge emitter and GaSb cell.

Depending on the coupling strength between these waveguide modes, radiative transfer can either be enhanced or suppressed. Stronger coupling occurs when the waveguide modes supported in the emitter and the PV cell overlap in both frequency and in-plane wavevector and vice-versa for weaker coupling. Therefore, by choosing appropriate thicknesses for the thin-film emitter and PV cell, it is possible to simultaneously enhance thermal radiation at wavelengths above the GaSb band gap and suppress thermal radiation at wavelengths below the band gap as shown in Fig. 2b. Between wavelengths of 1.1 μm and 1.6 μm, a strong band exists which indicates similar waveguide modes are supported in both the emitter and PV cell resulting in strong coupling and thus large heat flux. However, at wavelengths shorter than 1.1 μm and wavelengths longer than 1.6 μm several weaker bands can be observed. This is due to the difference in thicknesses between the emitter and PV cell where the more numerous waveguide modes supported in the emitter can only weakly couple to the few modes that are above the cut-off frequencies supported by the PV cell in this wavelength range.

Although this mismatch in waveguide modes also reduces the relative coupling strength for above band-gap modes, the suppression of sub-band gap radiative transfer is more crucial in improving TPV performance for a Ge emitter and GaSb cell combination. Furthermore, since GaSb and Ge are both semiconductors with similar band gaps, both materials exhibit an increase in the imaginary component of the permittivity at wavelengths above the band gap. This results in the broadening of the waveguide modes which allows the increased radiative energy transfer from the emitter to the PV cell. Although the mechanism of tailoring spectral thermal emission from low-dimensional structures is general and applicable to any material combination of the emitter and the PV cell, the optimum thermal well thicknesses are specific to a particular combination of materials and operating temperatures.

For comparison with previously proposed near-field TPV designs, a bulk W emitter was also calculated. Figure 2c and 2d show the corresponding transmission functions for a bulk and a thin-film GaSb cell, respectively. Once again, the transmission function for the bulk system has a broad spectrum similar to Fig. 2a. In the thin-film case, we can still observe the formation of distinct bands. When compared to Fig. 2b, fewer bands are observed which indicates that these bands are solely due to the waveguide modes in the PV cell. Similar to Fig. 2b, a distinct band can once again be observed from 1.1 μm to 1.6 μm. The peak at 950 nm is due to the surface plasmon polariton modes supported in the W emitter.

Figures 3a and 3b show the spectral heat flux for a Ge emitter and W emitter, respectively, again for the case that the back reflector is a perfect metal. For both cases, the emitter temperature was assumed to be 1000 K and the gap separation assumed to be 100 nm. In Fig. 3a, a progressive decrease in the thickness of the emitter and the PV cell results in strong spectral shaping of the heat flux where long wavelength thermal radiation is significantly suppressed and short wavelength thermal radiation is enhanced resulting in thermal radiation higher than the blackbody limit at the same temperature. This behavior is expected as photonic confinement effects only occur when the thickness of the thin-films are comparable to or smaller than the wavelength of IR radiation. In this regime, the cut-off frequencies of the waveguides modes will blue shift as the thickness decreases. Therefore, long wavelength radiation is inherently more sensitive to variations in thickness. By comparison, the short wavelength range is relatively insensitive to variations in thickness and from Fig. 3a it can be observed in that even for thicknesses of 5 μm, the spectral heat flux overlaps with the bulk structure at wavelengths shorter than 1.5 μm.

For the case of a bulk W emitter, Fig. 3b shows that as the thickness of the GaSb cell decreases, long wavelength thermal radiation is still suppressed while short wavelength thermal radiation is again enhanced beyond the blackbody limit. This suggests that despite the broad emission of SPP modes, the thermal well effect can still be utilized to dramatically improve the spectral selectivity by simply making the PV cell thin in order to suppress absorption at longer wavelengths.

At this point, it is worth mentioning that Ge at high temperatures will also exhibit a broader emission spectrum due to a combination of band gap shrinkage and free carrier emission. For the sake of demonstrating the thermal well concept, this effect was not included in the results plotted in Figs. 2 and 3. Although neglecting this effect may appear to be an oversimplification, the results for the bulk W emitter indicate that despite the presence of long wavelength emission, the inability of the GaSb cell to absorb thermal radiation in this wavelength range will ensure that the spectral selectivity of the TPV system is maintained. To confirm this assumption, additional calculations were performed in which the extinction coefficient of Ge was artificially raised to emulate the effects of high temperature and the results show that the spectral selectivity was still improved via the thermal well effect (see Supplementary Information for further details).

**Improving TPV System Performance using Thin-Film PV Cells.** To assess the TPV conversion efficiency of the TPV system we combine the radiative heat transfer model with an electrical model that considers the recombination of charge carriers in the PV cell (see Methods). In this model, the PV cell is assumed to be a diode which consists of a p-type quasi-neutral region, n-type quasi-neutral region, and a space charge region as shown in Fig. 1b. In order to assess the impact of the emitter and cell morphology on TPV system performance, a comparison can be made between the efficiencies for the thin-film structures with a maximum efficiency calculated from the Shockley Queisser formulation which assumes no non-radiative recombination processes in the PV cell and only far-field radiative transfer occurs.[56] If the emitter is a blackbody at a temperature of $T_H$ = 1000 K and the PV cell is at temperature of $T_C$ = 300 K, the Shockley Queisser formulation predicts the maximum efficiency achievable for a GaSb cell is 1.5%. For an emitter temperature of $T_H$ = 2000 K, the efficiency increases to 20.7%.

Using the calculated radiative power spectra and equations (3-9) (see Methods), the predicted efficiencies for the thin-film structure, which again includes non-radiative bulk and surface recombination losses, are shown in Fig. 4a as a function of temperature assuming a gap separation of 100 nm. At an emitter temperature of $T_H$ = 1000 K, the predicted efficiencies in the bulk limit is 0.38% for a Ge emitter and 1.9% for a W emitter due to the broad spectrum of radiative heat transfer shown in Fig. 3. By applying the thermal well effect and reducing the dimensionality of the emitter and PV cell, the energy conversion efficiency can be improved dramatically. For the case where perfect metals are used for both back reflectors, the efficiency reaches 38.7% for a thin-film Ge emitter and 28.7% for a bulk W emitter, which is more than an order of magnitude higher than the bulk limit. These predicted efficiencies also exceed past TPV efficiency records of 22% for GaSb cells while using a substantially lower emitter temperature of 1000 K compared to temperatures higher than 1500 K.[9,57]

To investigate this enhancement further, if the perfect metal back reflector on a thin-film Ge emitter is replaced with W, it can be observed that at low temperatures the efficiency is lower compared to the case of a perfect metal due to broadening of thermal emission at longer wavelengths. However, at higher temperatures, shorter wavelength modes are preferentially excited as the Planck energy oscillator function blue shifts. As a result, the short wavelength SPP mode supported in W contributes more to radiative transfer compensating for the long wavelength emission. In conjunction with thermal emission from the thin-film Ge emitter, this combination can actually lead to an even higher efficiency of 39.4% again at $T_H$ = 1000 K.

In order to provide a prediction of efficiency for a more realistic system, the back reflector on the PV cell is replaced by a Ag substrate separated from the PV cell using a transparent $MgF_2$ spacer layer. In this case, the efficiency decreases to 20.8% for a thin-film Ge emitter supported by a W substrate and 14.5% for a bulk W emitter at $T_H$ = 1000 K. This reduction in performance is due to the use of W and Ag which not only supports broad SPP modes that can couple in the near-field regime, but also exhibit intrinsic parasitic absorption and emission at longer wavelengths due to the imperfect nature of these materials as back reflectors. Although the $MgF_2$ spacer layer reduces SPP mode coupling by increasing the distance between the back reflectors, long wavelength absorption and emission still inhibit the predicted performance.

Despite this reduction in performance, the predicted efficiencies for all cases not only exceed the bulk limit, but also the Shockley Queisser limit for a blackbody emitter by several orders of magnitude. Figure 4b shows the predicted efficiency normalized to the efficiency from the Shockley

Queisser formulation assuming the emitter is a blackbody and the PV cell has a band gap of 0.726 eV. For certain cases, the enhancement reaches its maximum at intermediate temperatures. This is due to the more sensitive nature of thermal emission in the near-field regime compared to blackbody emission where lower temperatures will red shift the population of radiating modes resulting in a more rapid decrease in thermal emission at wavelengths where strong evanescent coupling occurs. The data presented in Figs. 3a and 3b demonstrate that the use of the thermal well effect can enable operation of near-field TPV systems with efficiencies higher than 30% at moderate temperatures of about 800 K.

Due to the inherent practical difficulties associated with utilizing the near-field regime in a TPV system, the conversion efficiency was also calculated as a function of the gap separation assuming an emitter temperature of $T_H$ = 1000 K as shown in Fig. 4c. For all cases, the efficiency saturates at gap separations larger than 5 μm which corresponds to the far-field limit. In this limit, for the case where both back reflectors are perfect metals, the efficiency decreases to 31.5% for a thin-film Ge emitter and 18.9% for a bulk W emitter. For the case of a thin-film Ge emitter and a W back reflector, the efficiency decreases to 34.6%. This reduction in performance is expected as the enhancement by near-field coupling is no longer utilized. However, despite this reduction, the predicted efficiencies still clearly exceed the Shockley Queisser limit for a blackbody emitter. The oscillatory behavior of efficiency at intermediate gap separations is due to the vacuum gap behaving like a waveguide.

In the case where a Ag back reflector and a $MgF_2$ spacer on the PV cell are used, the efficiency decreases more significantly to 4.9% for a thin-film Ge emitter supported by a W substrate and 7.4% for a bulk W emitter. Again this can be attributed to parasitic emission and absorption in W and Ag which lead to significant radiative transfer at longer wavelengths which cannot be used for power generation.

## Discussion

The enhancement in TPV system efficiency, as demonstrated in Fig. 4, can be attributed to improvements to both the spectral selectivity of radiative transport via the thermal well effect, as was shown in Fig. 3, and a reduction in bulk recombination losses for a thin-film PV cell which results in a lower saturation current. It should be emphasized that although the thermal well effect was proposed as a way to manipulate the photon DOS in the near-field regime, it can also be used to improve the spectral selectivity in the far-field limit as evidenced by the high efficiencies predicted for the cases that use a perfect metal as a back reflector in Fig. 4c. Again this can be attributed to the creation of quantized waveguide modes in the emitter and the PV cell. Although resonant enhancement for high photon energies larger than the band gap of the PV cell will be weaker due to the lack of near-field coupling, the suppression of emission and absorption below the cut-off frequency of these waveguide modes still ensures strong spectral selectivity in the far-field regime.

In regards to the electrical performance of the GaSb cell, the thickness of the space charge region formed in a GaSb cell is estimated to be 135 nm based on equation (9) (see Methods) and the properties of a typical GaSb cell. For the various cases presented, the PV thin-film thicknesses are either comparable to or smaller than the predicted size of the space charge region. This implies that the p-type and n-type quasi-neutral regions in the PV cell are so thin that bulk recombination losses are negligible. Therefore, according to the simple diode model, the saturation current based on equation (8) is dominated by surface recombination. If the PV cell is well passivated, surface recombination losses will be comparatively smaller than bulk recombination losses, thus a thin-film PV cell will always yield a lower saturation current compared to a bulk PV cell. This in turn will result in a higher open-circuit voltage increasing the efficiency.

To show the impact of reducing bulk recombination losses on the PV performance, the predicted efficiency was also calculated using the Shockley-Queisser formulation assuming the same radiative power density for each case (see Supplementary Information). When both back reflectors are perfect metals, the predicted efficiencies at an emitter temperature of $T_H$ = 1000 K and a gap separation of g = 100 nm are 46.4% and 34.5% for a thin-film Ge emitter and a bulk W emitter, respectively. For a thin-film Ge emitter supported by a W back reflector, the efficiency is 46.7%. And finally, for the PV cell back reflector composed of a Ag mirror and a $MgF_2$ spacer, the efficiency is 25.2% and 17.4% for a thin-film Ge emitter and a bulk W emitter, respectively. In all cases, the efficiencies calculated using the more realistic electrical model approach the performance predicted by the ideal Shockley Queisser formulation. Therefore it is clear that by using thin-film PV cells, the corresponding reduction in bulk recombination losses can dramatically improve the electrical performance of the device.

It was also observed that by combining different emitting mechanisms, namely thin-film waveguide modes in Ge and SPP modes in W, it is possible at high temperatures to exceed the efficiency predicted using a perfect metal as the back reflector on the emitter. This suggests that there exists some flexibility in the design of the emitter and with an optimal material combination for the thin-film and substrate, even higher energy conversion efficiencies can be obtained compared to the predictions in this study. The key however is that a perfect mirror needs to be used as a back reflector for the PV cell to eliminate parasitic absorption and emission at photon energies smaller than the band gap of the PV cell. As shown, once the perfect metal on the PV cell is replaced with Ag and a $MgF_2$ spacer, the performance reduces significantly. Although no natural materials behave like a perfect metal, it is possible to design artificial photonic structures that can emulate the behavior of a perfect metal within a certain wavelength range. For example, a distributed Bragg reflector, which is commonly used as a high quality mirror, supports a photonic band gap which can be positioned at photon energies just below the electronic band gap of the PV cell.[58,59] In fact, the concept of incorporating a Bragg reflector in a PV module is not new and has been used in the past as a way to recycle photons in single junction PV systems.[60–62]

To improve the performance of the TPV system even further, it should be stressed that the choice of the materials in this study is not necessarily the most optimal. For example, the band gap of the GaSb cell is relatively large compared to the low emitter temperatures thus inhibiting the portion of thermal radiation that can be used for power generation. Smaller band gap ternary and quaternary PV cells, such as InGaAsSb or InGaSb devices, can extend the range of useful photons for power generation to longer wavelengths resulting in both a higher efficiency and a higher electrical power density at low emitter temperatures. In fact, given that the optimal PV cell thickness in this study is significantly thinner than typical epitaxially grown ternary and quaternary devices, it is likely that TPV systems utilizing these materials will benefit from the thermal well effect.[14,63–65] Furthermore, the inherent simplicity of reducing the dimensionality of the emitter and the PV cell allows conventional components such as filters or antireflection coatings to be easily incorporated which can improve the spectral selectivity even further.

By improving the spectral selectivity and reducing the saturation current, the studied thermal well effect is theoretically predicted to enhance the energy conversion efficiency of TPV systems by more than an order of magnitude compared to both the bulk limit and the Shockley Queisser limit for a blackbody emitter at the same temperature. For a thin-film Ge emitter and a thin-film GaSb PV cell supported by perfect metals, the TPV energy conversion efficiency was predicted to be as high as 38.7% at an emitter temperature of 1000 K and gap separation of 100 nm due to only the thermal well effect. In the far-field limit, this efficiency decreases to 31.5%; however, this is still significantly higher than the Shockley Queisser limit even for a blackbody emitter at a temperature of 2000 K. This is in stark contrast to past studies that utilized SPP modes to improve TPV systems as the efficiency enhancement is limited to the near-field regime which is challenging to realize in a practical system.

Overall, thermal well TPV systems can provide higher TPV efficiency at lower emitter temperatures much like quantum well lasers which feature lower pumping thresholds than conventional diode lasers while being exceedingly more efficient or quantum well thermoelectric generators, which have thermoelectric figures of merit that are much higher than bulk systems.[66] By introducing morphological effects into a TPV system, not only can greater flexibility in engineering radiative heat transfer between the emitter and the PV cell be achieved, but also even simple, easily fabricated structures can have a profound effect on the overall performance of the TPV system.

## Methods

**Rytov Theory.** We model the energy exchange between the thermal emitter and the PV cell using a rigorous analytical electromagnetic formulation based on Rytov theory, in which thermally emitting bodies are modelled as a volume of fluctuating dipole sources whose amplitude is determined by the fluctuation dissipation theorem.[47–49,67–71] Dyadic Green's functions are then used to calculate the Poynting vector at a position relative to the emitting body. In this manner, the rate of radiative energy transfer, or heat flux, can be determined from one medium to another. It should be noted that this formalism is valid in both the near-field and far-field regimes. The general form of the heat flux is as follows,

$$q_{mn} = \frac{\theta(\omega, T_m)}{\pi^2} \cdot \underbrace{2\pi k_v^2 \cdot \text{Re}\left\{i\varepsilon'' \int dV' \left[G_{E,xn}(\bar{r},\bar{r}',\omega) G_{H,yn}^*(\bar{r},\bar{r}',\omega) - G_{E,yn}(\bar{r},\bar{r}',\omega) G_{H,xn}^*(\bar{r},\bar{r}',\omega)\right]\right\}}_{G_T} \quad (1)$$

where $q_{mn}$ is the heat flux from medium m to n, $\theta$ is the energy for a Planck oscillator, and $G_E/G_H$ are the dyadic Green's functions for the electric and magnetic fields.[72] The term $G_T$ is defined as the transmission function and is dependent on both frequency and the in-plane wavevector of photons participating in the radiative heat exchange. The transmission function represents the available radiative channels for energy transport between the emitter and PV cell. Therefore, any modification of the photon DOS imposed on the system by changing its geometry and/or material will manifest itself directly in the transmission function.

In order to utilize equation (1) to find the heat flux, specific solutions to the dyadic Green's functions must be found for a particular geometry. For the system defined in Fig. 1, the analytical solutions can be found by using a combination of the transfer matrix and scattering matrix methods.[73] Further details of the formulation can be found in the supplementary information. With this formulation, the heat flux into a particular layer is determined as the difference between the incoming and outgoing Poynting vectors for that layer. Therefore, the net heat flux into the PV cell, $q_\omega$, can be determined as follows,

$$q_\omega = q_{03} + q_{13} - q_{31} - q_{30} \quad (2)$$

where $q_{03}$ is the heat flux from the emitter back reflector to the PV cell, $q_{13}$ is the heat flux from the thin-film emitter to the PV cell, $q_{31}$ is the heat flux from the PV cell to the thin-film emitter, and $q_{30}$ is the heat flux from the PV cell to the emitter back reflector. By using equations (1) and (2), the radiative power spectrum absorbed by the PV cell can be calculated for a particular structure and set of materials.

**TPV System Efficiency Calculation.** In order to provide a more realistic estimation of the TPV system performance compared to the more idealized Shockley Queisser formulation, we incorporated specific material recombination lifetimes to account for both radiative and non-radiative recombination losses. Following the conventional definition for the photovoltaic energy conversion efficiency, we define the efficiency of our TPV system as,[56,74,75]

$$\eta = \frac{P_E}{P_R} = \frac{I_{SC} \cdot V_{OC} \cdot FF}{P_R} \quad (3)$$

where $\eta$ is the TPV energy conversion efficiency, $P_E$ is the electrical power density, $P_R$ is the radiative power density, FF is the fill factor, $I_{SC}$ is the short-circuit current, and $V_{OC}$ is the open-circuit voltage. In this model, the short-circuit current is assumed to be approximately equal to the photo-generated current, $I_{PH}$. The radiative power density is the net radiative heat transfer between the emitter and the PV cell integrated over all frequencies,

$$P_R = \int_0^\infty (q_{03} + q_{04} + q_{13} + q_{14} - q_{31} - q_{30} - q_{41} - q_{40}) d\omega \tag{4}$$

where $q_{04}$ is the heat flux from emitter back reflector to the PV cell back reflector, $q_{14}$ is the heat flux from the thin-film emitter to the PV cell back reflector, $q_{41}$ is the heat flux from the PV cell back reflector to the thin-film emitter, and $q_{40}$ is the heat flux from the PV cell back reflector to the emitter back reflector. To calculate the short-circuit current, $I_{SC}$, the power spectrum is integrated for photon energies larger than the band gap of the PV cell. By assuming each absorbed photon produces one electron, the short-circuit current can be obtained as follows,

$$I_{SC} \approx I_{PH} = \int_{\frac{E_g}{\hbar}}^\infty \frac{e}{\hbar\omega} \cdot q_\omega d\omega \tag{5}$$

The open-circuit voltage is determined by taking the limit of zero current in the general diode equation. This can be expressed as,

$$V_{OC} = \frac{k_b T_C}{e} \cdot \ln\left(\frac{I_{SC}}{I_0} + 1\right) \tag{6}$$

where $I_0$ is the saturation current, $k_b$ is the Boltzmann constant, and e is the electron charge. The prefactor $(k_b T_C/e)$ represents the thermally induced potential. Finally, the fill factor is defined as the maximum electrical power output from the PV cell normalized by the product of the short-circuit current and open-circuit voltage,

$$FF = \frac{\max(I \cdot V)}{I_{SC} \cdot V_{OC}} \tag{7}$$

To determine the saturation current, an analytical expression can be obtained by considering the diffusion of carriers in the PN junction under zero external bias. For a finite size PN junction, this expression is as follows,[74]

$$I_0 = \frac{eD_n n_i^2}{L_n N_p} \cdot F_n + \frac{eD_p n_i^2}{L_p N_n} \cdot F_p \; ; \; F_x = \frac{S_x \cdot \cosh(t_x/L_x) + D_x/L_x \cdot \sinh(t_x/L_x)}{D_x/L_x \cdot \cosh(t_x/L_x) + S_x \cdot \sinh(t_x/L_x)} \tag{8}$$

where is $n_i$ the intrinsic carrier concentration, $N_n$ is the n-type dopant concentration, $N_p$ is the p-type dopant concentration, $D_x$ is the carrier diffusivity, $S_x$ is the surface recombination velocity, $t_x$ is the thickness of the quasi-neutral regions, and $L_x$ is the diffusion length. The subscripts n and p denote electrons and holes, respectively. The diffusion length can be defined in terms of a total recombination lifetime, $L = \sqrt{D \cdot \tau}$. The recombination lifetime is an inverse summation over all recombination mechanisms including radiative recombination, Shockley-Read-Hall recombination, and Auger recombination.[74] The thickness of the p-type and n-type quasi-neutral regions are assumed equal and are obtained by taking the difference between the total thickness of the PV cell and the thickness of the space charge region which is expressed as,[76]

$$t_{SCR} = \sqrt{\frac{2k_b T_C \varepsilon}{e^2} \cdot \ln\left(\frac{N_n N_p}{n_i^2}\right) \cdot \left(\frac{1}{N_n} + \frac{1}{N_p}\right)} \tag{9}$$

where ε is the permittivity of the PV cell taken in the long wavelength limit. Since the PV cell thickness is chosen independently in the radiative heat transfer model, if the thickness of the PV cell is smaller than the space charge region, the thicknesses of the p-type and n-type quasi-neutral regions are assumed to be negligible so that the PV cell consists entirely of a fully depleted region. In this limit, surface recombination becomes the dominant loss mechanism.

The electrical properties of the GaSb cell were obtained from literature.[77] It is assumed the GaSb cell is at a uniform temperature of 300K. The intrinsic carrier concentration, $n_i$, is assumed to be $4.3 \cdot 10^{12}$ cm$^{-3}$. The electron and hole carrier concentrations are equal to $N_n = N_p = 10^{17}$ cm$^{-3}$. The recombination lifetimes are $\tau_R$ = 40 ns, $\tau_{SHR}$ = 10 ns, and $\tau_{Au}$ = 20 μs for radiative recombination, Shockley-Read-Hall recombination, and Auger recombination, respectively. The carrier diffusivities are $D_n$ = 129 cm$^2$/s and $D_p$ = 39 cm$^2$/s for electrons and holes, respectively. The surface recombination velocity is chosen to be $S_n = S_p$ = 100 cm/s in accordance to previous studies.[53]

**References**

1. Kolm, H. H. *Solar-battery power source*. Tech. Rep., MIT Lincoln Laboratory, p.13 (1956).
2. Aigrain, P. Thermophotovoltaic conversion of radiant energy; unpublished lecture series at MIT. (1956).
3. Swanson, R. M. Recent developments in Thermophotovoltaic Conversion. *Proc. Int. Electron. Devices Meet.* 186–189 (1980).
4. Wedlock, B. D. Thermo-photo-voltaic energy conversion. *Proc. IEEE* **51,** 694–698 (1963).
5. Harder, N.-P. & Würfel, P. Theoretical limits of thermophotovoltaic solar energy conversion. *Semicond. Sci. Technol.* **18,** S151–S157 (2003).
6. Rephaeli, E. & Fan, S. Absorber and emitter for solar thermo-photovoltaic systems to achieve efficiency exceeding the Shockley-Queisser limit. *Opt. Express* **17,** 15145–59 (2009).
7. Datas, A. & Algora, C. Global optimization of solar thermophotovoltaic systems. *Prog. Photovolt Res. Appl.* **21**, 1040–1055 (2013).
8. Zenker, M. *et al*. Efficiency and power density potential of combustion-driven thermophotovoltaic systems using GaSb photovoltaic cells. *IEEE Trans. Electron Devices* **48,** 367–376 (2001).
9. Fraas, L. M. *et al*. Thermophotovoltaic system configurations and spectral control. *Semicond. Sci. Technol.* **18,** S165–S173 (2003).
10. Catalano, A. Thermophotovoltaics: A new paradigm for power generation? *Renew. Energy* **8,** 495–499 (1996).
11. Wurfel, P. & Ruppel, W. Upper limit of thermophotovoltaic solar-energy conversion. *IEEE Trans. Electron Devices* **27,** 745–750 (1980).
12. Demichelis, F. & Minetti-Mezzetti, E. A solar thermophotovoltaic converter. *Sol. Cells* **1,** 395–403 (1980).
13. Coutts, T. J. Thermophotovoltaic principles, potential, and problems. *AIP Conf. Proc.* **404,** 217–234 (1997).
14. Basu, S., Chen, Y.-B. & Zhang, Z. M. Microscale radiation in thermophotovoltaic devices—A review. *Int. J. Energy Res.* **31,** 689–716 (2007).
15. Celanovic, I. *et al*. 1D and 2D Photonic Crystals for Thermophotovoltaic Applications. *Proc. SPIE* **5450**, 416–422 (2004).
16. Coutts, T. J. A review of progress in thermophotovoltaic generation of electricity. *Renew. Sustain. Energy Rev.* **3,** 77–184 (1999).
17. Coutts, T. J. & Ward, J. S. Thermophotovoltaic and photovoltaic conversion at high-flux densities. *IEEE Trans. Electron Devices* **46,** 2145–2153 (1999).
18. Durisch, W. & Bitnar, B. Novel thin film thermophotovoltaic system. *Sol. Energy Mater. Sol. Cells* **94,** 960–965 (2010).
19. Fraas, L. M. *et al*. Over 35-percent efficient GaAs/GaSb tandem solar cells. *IEEE Trans. Electron Devices* **37,** 443–449 (1990).
20. Nam, Y. *et al.* Solar thermophotovoltaic energy conversion systems with two-dimensional tantalum photonic crystal absorbers and emitters. *Sol. Energy Mater. Sol. Cells* **122,** 287–296 (2014).

21. Narayanaswamy, A. 1D Metallo-Dielectric Photonic Crystals as Selective Emitters for Thermophotovoltaic Applications. *AIP Conf. Proc.* **738,** 215–220 (2004).
22. Yeng, Y. X. *et al.* Performance analysis of experimentally viable photonic crystal enhanced thermophotovoltaic systems. *Opt. Express* **21 S6,** A1035–51 (2013).
23. Zhao, B. *et al.* Thermophotovoltaic emitters based on a two-dimensional grating/thin-film nanostructure. *Int. J. Heat Mass Transf.* **67,** 637–645 (2013).
24. Shen, Y. *et al.* Optical Broadband Angular Selectivity. *Science* **343 ,** 1499–1501 (2014).
25. Bermel, P. *et al.* Design and global optimization of high-efficiency thermophotovoltaic systems. *Opt. Express* **18 S3,** A314–34 (2010).
26. Chan, W. R. *et al.* Toward high-energy-density, high-efficiency, and moderate-temperature chip-scale thermophotovoltaics. *Proc. Natl. Acad. Sci.* **110,** 5309–14 (2013).
27. Datas, A. & Algora, C. Development and experimental evaluation of a complete solar thermophotovoltaic system. *Prog. Photovolt Res. Appl.* 1025–1039 (2013).
28. Lenert, A. *et al.* A nanophotonic solar thermophotovoltaic device. *Nat. Nanotechnol.* **9,** 126–30 (2014).
29. Wernsman, B. *et al.* Greater than 20% radiant heat conversion efficiency of a thermophotovoltaic radiator/module system using reflective spectral control. *IEEE Trans. Electron Devices* **51,** 512–515 (2004).
30. Colangelo, G., de Risi, A. & Laforgia, D. Experimental study of a burner with high temperature heat recovery system for TPV applications. *Energy Convers. Manag.* **47,** 1192–1206 (2006).
31. Crowley, C. J. Thermophotovoltaic Converter Performance for Radioisotope Power Systems. *AIP Conf. Proc.* **746,** 601–614 (2005).
32. Ferrari, C., Melino, F., Pinelli, M. & Spina, P. R. Thermophotovoltaic energy conversion: Analytical aspects, prototypes and experiences. *Appl. Energy* **113,** 1717–1730 (2014).
33. DiMatteo, R. S. Micron-gap ThermoPhotoVoltaics (MTPV). *AIP Conf. Proc.* **653,** 232–240 (2003).
34. DiMatteo, R. S. *et al.* Enhanced photogeneration of carriers in a semiconductor via coupling across a nonisothermal nanoscale vacuum gap. *Appl. Phys. Lett.* **79,** 1894 (2001).
35. Laroche, M., Carminati, R. & Greffet, J.-J. Near-field thermophotovoltaic energy conversion. *J. Appl. Phys.* **100,** 063704 (2006).
36. Narayanaswamy, A. & Chen, G. Surface modes for near field thermophotovoltaics. *Appl. Phys. Lett.* **82,** 3544 (2003).
37. Niv, A. *et al*. Near-Field Electromagnetic Theory for Thin Solar Cells. *Phys. Rev. Lett.* **109,** 138701 (2012).
38. Pan, J. L., Choy, H. K. H. & Fonstad, C. G. Very large radiative transfer over small distances from a black body for thermophotovoltaic applications. *IEEE Trans. Electron Devices* **47,** 241–249 (2000).
39. Whale, M. D. & Cravalho, E. G. Modeling and performance of microscale thermophotovoltaic energy conversion devices. *IEEE Trans. Energy Convers.* **17,** 130–142 (2002).
40. Ilic, O. *et al*. Overcoming the black body limit in plasmonic and graphene near-field thermophotovoltaic systems. *Opt. Express* **20,** A366 (2012).
41. Dingle, R., Wiegmann, W. & Henry, C. Quantum States of Confined Carriers in Very Thin Al_{x}Ga_{1-x}As-GaAs-Al_{x}Ga_{1-x}As Heterostructures. *Phys. Rev. Lett.* **33,** 827–830 (1974).
42. Dupuis, R. D. *et al*. Continuous 300 °K laser operation of single-quantum-well AlxGa1−xAs-GaAs heterostructure diodes grown by metalorganic chemical vapor deposition. *Appl. Phys. Lett.* **34,** 265 (1979).
43. Boriskina, S. V., Ghasemi, H. & Chen, G. Plasmonic materials for energy: From physics to applications. *Mater. Today* **16,** 375–386 (2013).
44. Popovic, Z. & Popovic, B. D. *Introductory Electromagnetics*. (Prentice Hall, 1999).

45. Boriskina, S. V. *et al*. Highly efficient full-vectorial integral equation solution for the bound, leaky, and complex modes of dielectric waveguides. *IEEE J. Sel. Top. Quantum Electron.* **8,** 1225–1232 (2002).
46. Boriskina, S. V. & Nosich, A. I. Radiation and absorption losses of the whispering-gallery-mode dielectric resonators excited by a dielectric waveguide. *IEEE Trans. Microw. Theory Tech.* **47,** 224–231 (1999).
47. Rytov, S. M., Kravtsov, Y. A. & Tatarskii, V. I. *Principles of Statistical Radiophysics 3: Elements of Random Fields*. (Springer-Verlag Berlin Heidelberg, 1989).
48. Shen, S., Narayanaswamy, A. & Chen, G. Surface phonon polaritons mediated energy transfer between nanoscale gaps. *Nano Lett.* **9,** 2909–2913 (2009).
49. Francoeur, M., Mengüç, M. P. & Vaillon, R. Spectral tuning of near-field radiative heat flux between two thin silicon carbide films. *J. Phys. D. Appl. Phys.* **43,** 075501 (2010).
50. Bett, A. W. & Sulima, O. V. GaSb photovoltaic cells for applications in TPV generators. *Semicond. Sci. Technol.* **18,** S184–S190 (2003).
51. Sulima, O. & Bett, A. W. Fabrication and simulation of GaSb thermophotovoltaic cells. *Sol. Energy Mater. Sol. Cells* **66,** 533–540 (2001).
52. Algora, C. Modelling And Manufacturing GaSb TPV Converters. *AIP Conf. Proc.* **653,** 452–461 (2003).
53. Fraas, L. M. *et al.* Fundamental characterization studies of GaSb solar cells. *Conf. Rec. Twenty-Second IEEE Photovolt. Spec. Conf.* (1991).
54. Stelmakh, V. *et al.* Evolution of sputtered tungsten coatings at high temperature. *J. Vac. Sci. Technol. A Vacuum, Surfaces, Film.* **31,** 061505 (2013).
55. Palik, E. D. *Handbook of optical constants of solids*. (Academic Press, 1997).
56. Shockley, W. & Queisser, H. J. Detailed Balance Limit of Efficiency of p-n Junction Solar Cells. *J. Appl. Phys.* **32,** 510 (1961).
57. Fraas, L. M., Avery, J. E. & Huang, H. X. Thermophotovoltaic furnace–generator for the home using low bandgap GaSb cells. *Semicond. Sci. Technol.* **18,** S247–S253 (2003).
58. Brudieu, B. *et al.* Sol-Gel Route Toward Efficient and Robust Distributed Bragg Reflectors for Light Management Applications. *Adv. Opt. Mater.* **2**, **11**, 1105-1112 (2014).
59. Lin, A. *et al*. Aperiodic and randomized dielectric mirrors: alternatives to metallic back reflectors for solar cells. *Opt. Express* **22 S3,** A880–94 (2014).
60. Bermel, P. *et al*. Improving thin-film crystalline silicon solar cell efficiencies with photonic crystals. *Opt. Express* **15,** 16986 (2007).
61. Isabella, O. *et al*. Design and application of dielectric distributed Bragg back reflector in thin-film silicon solar cells. *J. Non. Cryst. Solids* **358,** 2295–2298 (2012).
62. Sheng, X. *et al.* Integration of Self-Assembled Porous Alumina and Distributed Bragg Reflector for Light Trapping in Si Photovoltaic Devices. *IEEE Photonics Technol. Lett.* **22,** 1394–1396 (2010).
63. Hitchcock, C. *et al.* Ternary and quaternary antimonide devices for thermophotovoltaic applications. *J. Cryst. Growth* **195,** 363–372 (1998).
64. Mauk, M. G. & Andreev, V. M. GaSb-related materials for TPV cells. *Semicond. Sci. Technol.* **18,** S191–S201 (2003).
65. Gonzalez-Cuevas *et al*. Modeling of the temperature-dependent spectral response of In(1-x)Ga(x)Sb infrared photodetectors. *Opt. Eng.* **45,** 044001 (2006).
66. Hicks, L. & Dresselhaus, M. Effect of quantum-well structures on the thermoelectric figure of merit. *Phys. Rev. B* **47,** 12727–12731 (1993).
67. Otey, C. & Fan, S. Numerically exact calculation of electromagnetic heat transfer between a dielectric sphere and plate. *Phys. Rev. B* **84,** 245431 (2011).
68. Mulet, J.-P. *et al*. Enhanced Radiative Heat Transfer At Nanometric Distances. *Microscale Thermophys. Eng.* **6,** 209–222 (2002).

69. Lee, B. J., Park, K. & Zhang, Z. M. Energy pathways in nanoscale thermal radiation. *Appl. Phys. Lett.* **91,** 153101 (2007).
70. Guo, Y. *et al*. Thermal excitation of plasmons for near-field thermophotovoltaics. *Appl. Phys. Lett.* **105,** 073903 (2014).
71. Ben-Abdallah, P. *et al*. Near-field heat transfer mediated by surface wave hybridization between two films. *J. Appl. Phys.* **106,** 044306 (2009).
72. Narayanaswamy, A. & Chen, G. Dyadic Green's functions and electromagnetic local density of states. *J. Quant. Spectrosc. Radiat. Transf.* **111,** 1877–1884 (2010).
73. Francoeur, M., Pinar Mengüç, M. & Vaillon, R. Solution of near-field thermal radiation in one-dimensional layered media using dyadic Green's functions and the scattering matrix method. *J. Quant. Spectrosc. Radiat. Transf.* **110,** 2002–2018 (2009).
74. Green, M. *Solar Cells- Operating Principles, Technology and System Applications*. Englewood Cliffs, NJ, Prentice-Hall, Inc. (1982)*.*
75. Laroche, M., Carminati, R. & Greffet, J.-J. Near-field thermophotovoltaic energy conversion. *J. Appl. Phys.* **100,** 063704 (2006).
76. Williams, B. W. *Power Electronics: Devices, Drivers, Applications and Passive Components*. (McGraw-Hill, 1992).
77. Rosencher, E., Vinter, B. & Piva, P. G. *Optoelectronics*. (Cambridge University Press, 2002).

## Acknowledgements
The authors would like to thank P. Sambegoro, M. Branham, A. Lenert, and D. Bierman for helpful discussions. This work was supported by DOE-BES Award No. DE-FG02-02ER45977 (for the thermal well concept) and by the 'Solid State Solar-Thermal Energy Conversion Center (S3TEC)', funded by the US Department of Energy, Office of Science, and Office of Basic Energy, Award No. DE-SC0001299/DE-FG02-09ER46577 (for thermophotovoltaic applications).

## Author contributions
S.V.B. conceived the original idea. J.K.T., Y.H., and S.V.B. developed the radiative heat transfer model. J.K.T., W.-C.H. developed the electronic model for the PV cell. G.C. supervised the research. J.K.T., S.V.B., and G.C. wrote the paper. All authors contributed to scientific discussions and comments on the manuscript.

## Additional information
**Competing financial interests:** The authors declare no competing financial interests.

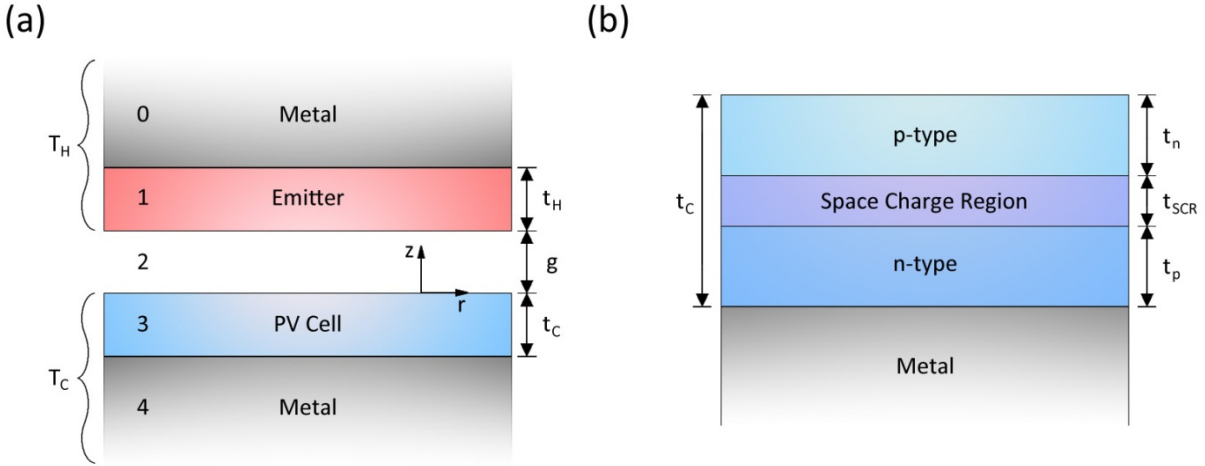

Figure 1: Schematics of the thin-film system studied for the radiative heat transfer model and the PV cell electrical model. (a) The TPV system consists of a thin-film emitter and a thin-film PV cell. Both films are placed onto semi-infinite back reflectors. (b) A magnified view of the PV cell detailing the structure assumed for the electrical analysis. The PV cell consists of a p-type quasi-neutral region, a n-type quasi-neutral region, and a space charge region. In this study, the thicknesses of the p-type and n-type quasi-neutral regions are assumed equal.

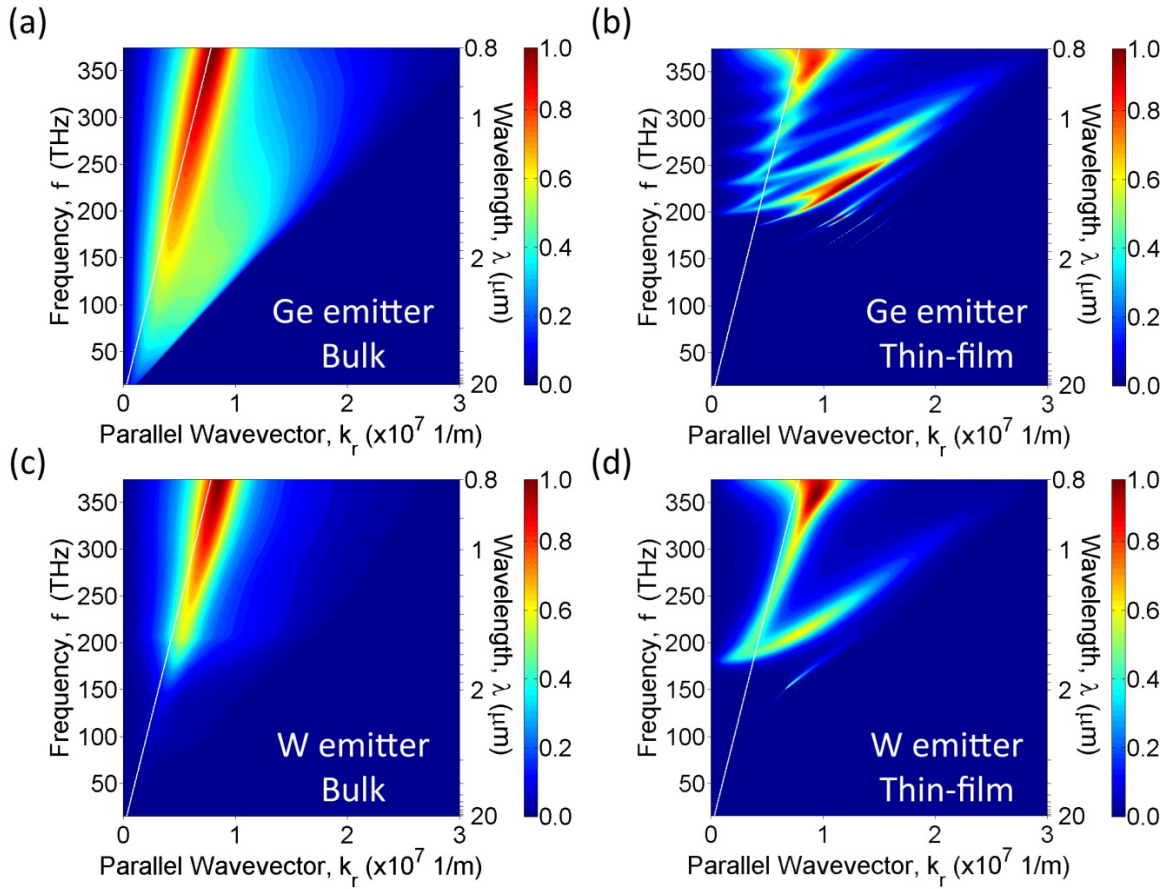

Figure 2: The normalized transmission function, $G_{NT}$, comparing a bulk system to a thin-film system for different emitter materials and a GaSb cell: (a) a bulk Ge emitter and bulk GaSb cell, (b) a thin-film Ge emitter and a thin-film GaSb cell. The thin-film thicknesses are $t_H$ = 860 nm and $t_C$ = 136 nm, (c) a bulk W emitter and a bulk GaSb cell, (d) a bulk W emitter and a thin-film GaSb cell. The thin-film thickness is $t_C$ = 134 nm. This comparison clearly showcases the effect of morphology on the trapped optical modes available for radiative transfer. The gap distance is assumed to be g = 100 nm. The light line is also plotted to differentiate propagating modes (above the light line) and evanescent modes (below the light line).

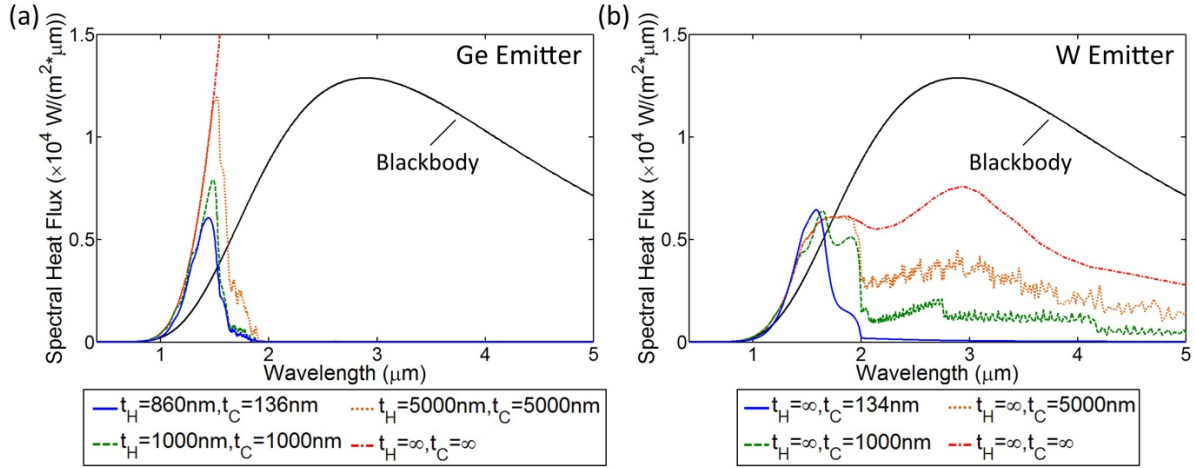

Figure 3: The spectral heat flux as a function of the emitter and the PV cell thicknesses assuming an emitter temperature $T_H$ = 1000 K and a gap distance g = 100 nm: (a) the spectral heat flux for a Ge emitter, (b) the spectral heat flux for a W emitter. By making the emitter and the PV cell thin, radiative energy transfer at wavelengths below the band gap is significantly suppressed.

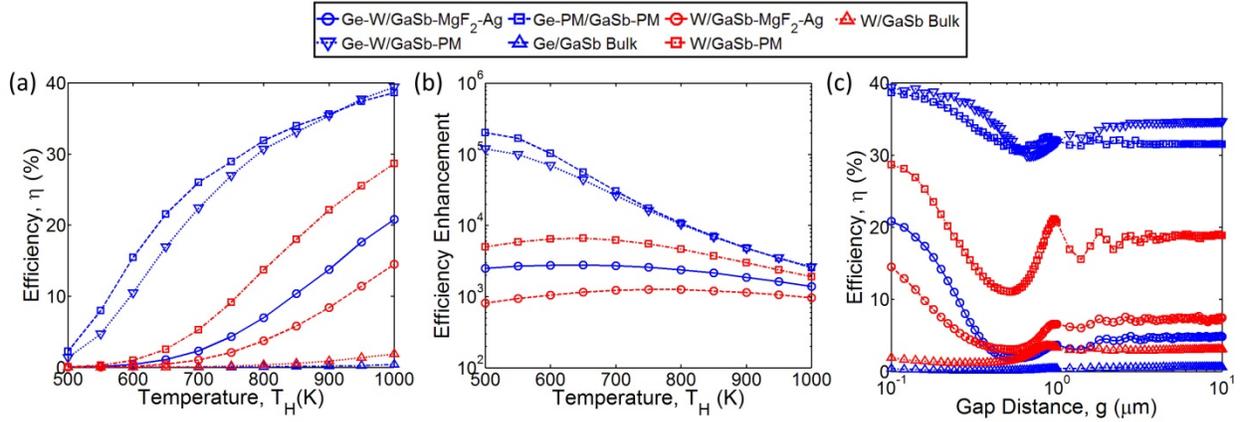

Figure 4: The energy conversion efficiency, η, from the emitter to the PV cell for several material combinations. (a) The efficiency as a function of temperature assuming a gap distance of g = 100 nm. (b) The predicted efficiencies from (a) normalized by efficiencies computed using the Shockley Queisser formulation for varying blackbody emitter temperatures. (c) The efficiency as a function of gap distance assuming an emitter temperature $T_H$ = 1000 K. The legend is identical to (a). The optimal thicknesses for the case of a Ge emitter on a W substrate and a GaSb cell on a Ag substrate with a $MgF_2$ spacer is $t_H$ = 119 nm, $t_C$ = 100 nm, and $t_S$ = 1.25 μm. The optimal thicknesses for a Ge emitter on a W substrate and a GaSb cell on a perfect metal are $t_H$ = 58 nm and $t_C$ = 94 nm. The optimal thicknesses for a bulk W emitter and a GaSb cell on a Ag substrate with a $MgF_2$ spacer is $t_C$ = 59 nm and $t_S$ = 750 nm.

# Supplementary Information

# Thin-film 'Thermal Well' Emitters and Absorbers for High-Efficiency Thermophotovoltaics

Jonathan K. Tong[1], Wei-Chun Hsu[1], Yi Huang[1], Svetlana V. Boriskina[1], and Gang Chen[1,*]

[1]*Department of Mechanical Engineering, Massachusetts Institute of Technology, Cambridge, MA 02139*

*Correspondence and requests for materials should be addressed to either S.V.B. or G.C. (emails: sborisk@mit.edu, gchen2@mit.edu)

**Near-field formalism for thin-film structures.** The thermal radiation model used in this study is a rigorous analytical formulation based on Rytov theory.[1–3] In this formulation, thermal emission is modelled electromagnetically as a volume of current sources. The spectral power density of these current sources is determined using the fluctuation-dissipation theorem and is defined as,

$$\langle J_i(r',\omega') J_j^*(r'',\omega'') \rangle = \frac{\omega \varepsilon_0 \varepsilon_r''(\omega)}{\pi} \theta(\omega,T) \cdot \delta(\omega'-\omega'') \delta(r'-r'') \delta_{ij} \tag{S1}$$

where $\varepsilon_r''$ is the imaginary component of permittivity, $\varepsilon_0$ is the vacuum permittivity, $\theta(\omega,T)$ is the mean energy of a Planck oscillator which generally includes the Bose-Einstein distribution and zero-point energy, r'/r'' are source positions in the emitting medium, $\delta(r'-r'')$ implies locality of the materials response, and $\delta_{ij}$ implies isotropy of the material. In this study, the system is assumed to consist of a 1D multilayer stack of thin-films. The derivation thus follows from previous studies.[3,4] The heat flux from one medium to another is determined by calculating the Poynting vector in the film of interest using the Dyadic Green's functions to connect the current sources to a position within the film. To simplify the derivation, a cylindrical coordinate system is assumed. The heat flux from medium *m* to medium *n* takes the following form,

$$q_{mn}(z) = \frac{k_v^2 \cdot \theta(\omega,T_m)}{\pi^2} \cdot \text{Re}\left\{ i\varepsilon'' \int dz' \int_0^\infty k_r dk_r \cdot \left[ G_{E,mn,rr}(z,z',\omega) G_{H,mn,\theta r}^*(z,z',\omega) + \ldots \right.\right.$$
$$\left.\left. \ldots + G_{E,mn,rz}(z,z',\omega) G_{H,mn,\theta z}^*(z,z',\omega) - G_{E,mn,\theta\theta}(z,z',\omega) G_{H,mn,r\theta}^*(z,z',\omega) \right] \right\} \tag{S2}$$

where $k_v$ is the vacuum wave vector, $k_r$ is the in-plane wavevector, $G_E$ is the Dyadic Green's function for the electric field, $G_H$ is the Dyadic Green's function for the magnetic field, z is a position in film *n*, and z' is the source position in film *m*.

In this system, four types of waves can be defined based on the direction of propagation in film *n* (+/- z) and the direction of propagation in the emitting film *m* (+/- z'). These four types of waves are incorporated into the expression of the Dyadic Green's functions. From (S2), the expressions for the product of Dyadic Green's functions for the electric and magnetic fields can be separated according to the polarization relative to the plane of the film. For TM polarization, the expression is as follows,

$$G_{E,mn,rr}G^*_{H,mn,\theta r}(k_r,z=z_n,\omega)+G_{E,mn,rz}G^*_{H,mn,\theta z}(k_r,z=z_n,\omega)=...$$

$$...=\frac{ik_{z,n}k^*_n}{8\mathrm{Re}(k_{z,m})\mathrm{Im}(k_{z,m})k_n|k_m|^2|k_{z,m}|^2}\cdot...$$

$$...\cdot\left[\mathrm{Re}(k_{z,m})\left(e^{2\mathrm{Im}(k_{z,m})t_m}-1\right)\left(|k_{z,m}|^2+k_r^2\right)\cdot\left(-|A_n^{TM}|^2-A_n^{TM}B_n^{TM*}+A_n^{TM*}B_n^{TM}+|B_n^{TM}|^2\right)\cdot...\right.$$

$$...\cdot i\cdot\mathrm{Im}(k_{z,m})\left(e^{-i\cdot 2\mathrm{Re}(k_{z,m})t_m}-1\right)\left(|k_{z,m}|^2-k_r^2\right)\cdot\left(A_n^{TM}C_n^{TM*}+A_n^{TM}D_n^{TM*}-B_n^{TM}C_n^{TM*}-B_n^{TM}D_n^{TM*}\right)\cdot...$$

$$...\cdot i\cdot\mathrm{Im}(k_{z,m})\left(1-e^{i\cdot 2\mathrm{Re}(k_{z,m})t_m}\right)\left(|k_{z,m}|^2-k_r^2\right)\cdot\left(A_n^{TM*}C_n^{TM}+B_n^{TM*}C_n^{TM}-A_n^{TM*}D_n^{TM}-B_n^{TM*}D_n^{TM}\right)\cdot...$$

$$\left....\cdot\mathrm{Re}(k_{z,m})\left(1-e^{-2\mathrm{Im}(k_{z,m})t_m}\right)\left(|k_{z,m}|^2+k_r^2\right)\cdot\left(-|C_n^{TM}|^2-C_n^{TM}D_n^{TM*}+C_n^{TM*}D_n^{TM}+|D_n^{TM}|^2\right)\right] \quad (S3)$$

where $t_m$ is the thickness of film m and the coefficients A, B, C, and D represent the amplitude of the four types of waves in the system. Similarly, for TE polarization,

$$G_{E,mn,\theta\theta}G^*_{H,mn,r\theta}(k_r,z=z_n,\omega)=...$$

$$...=\frac{ik^*_{z,n}}{8\mathrm{Re}(k_{z,m})\mathrm{Im}(k_{z,m})|k_{z,m}|^2}\cdot...$$

$$...\cdot\left[\mathrm{Re}(k_{z,m})\left(e^{2\mathrm{Im}(k_{z,m})t_m}-1\right)\cdot\left(|A_n^{TE}|^2-A_n^{TE}B_n^{TE*}+A_n^{TE*}B_n^{TE}-|B_n^{TE}|^2\right)\cdot...\right.$$

$$...\cdot i\cdot\mathrm{Im}(k_{z,m})\left(e^{-i\cdot 2\mathrm{Re}(k_{z,m})t_m}-1\right)\cdot\left(A_n^{TE}C_n^{TE*}-A_n^{TE}D_n^{TE*}+B_n^{TE}C_n^{TE*}-B_n^{TE}D_n^{TE*}\right)\cdot... \quad (S4)$$

$$...\cdot i\cdot\mathrm{Im}(k_{z,m})\left(1-e^{i\cdot 2\mathrm{Re}(k_{z,m})t_m}\right)\cdot\left(A_n^{TE*}C_n^{TE}-B_n^{TE*}C_n^{TE}+A_n^{TE*}D_n^{TE}-B_n^{TE*}D_n^{TE}\right)\cdot...$$

$$\left....\cdot\mathrm{Re}(k_{z,m})\left(1-e^{-2\mathrm{Im}(k_{z,m})t_m}\right)\cdot\left(|C_n^{TE}|^2-C_n^{TE}D_n^{TE*}+C_n^{TE*}D_n^{TE}-|D_n^{TE}|^2\right)\right]$$

A combination of the transfer matrix method and the scattering matrix method is used to recursively obtain analytical solutions for the coefficients A, B, C, and D. As an example, based on Fig. 1a of the main text, if the heat flux from medium 1 to medium 3 is desired, the coefficients take on the following form,

$$A_3^{TM,TE}=\frac{e^{ik_{z,1}t_H}e^{ik_{z,2}g}t_{21}^{TM,TE}t_{23}^{TM,TE}}{\left(1+r_{21}^{TM,TE}r_{10}^{TM,TE}e^{2ik_{z,1}t_H}\right)\left(1+r_{23}^{TM,TE}r_{34}^{TM,TE}e^{2ik_{z,3}t_C}\right)\left(1-r_{20}^{TM,TE}r_{24}^{TM,TE}e^{2ik_{z,2}g}\right)}\cdot\frac{k_{z,1}}{k_{z,2}}$$

$$B_3^{TM,TE}=r_{34}^{TM,TE}e^{2ik_{z,3}t_C}\cdot A_3^{TM,TE}$$

$$C_3^{TM,TE}=r_{34}^{TM,TE}\cdot A_3^{TM,TE} \quad (S5)$$

$$D_3^{TM,TE}=\left(r_{34}^{TM,TE}\right)^2 e^{2ik_{z,3}t_C}\cdot A_3^{TM,TE}$$

where r and t are the Fresnel reflection and transmission coefficients for either an interface or a slab. For TM and TE polarizations, respectively, these coefficients are,

$$r_{mn}^{TM} = \frac{\varepsilon_n k_{z,m} - \varepsilon_m k_{z,n}}{\varepsilon_n k_{z,m} + \varepsilon_m k_{z,n}}$$

$$t_{mn}^{TM} = \frac{2k_{z,m}\sqrt{\varepsilon_m}\sqrt{\varepsilon_n}}{\varepsilon_n k_{z,m} + \varepsilon_m k_{z,n}} \tag{S6}$$

$$r_{mn}^{TE} = \frac{k_{z,m} - k_{z,n}}{k_{z,m} + k_{z,n}}$$

$$t_{mn}^{TE} = \frac{2k_{z,m}}{k_{z,m} + k_{z,n}} \tag{S7}$$

For a thin slab, the analytical expression for the reflection and transmission coefficients are,

$$r_{mo}^{TM,TE} = \frac{r_{mn}^{TM,TE} + r_{no}^{TM,TE} e^{2ik_{z,n}t_n}}{1 + r_{mn}^{TM,TE} r_{no}^{TM,TE} e^{2ik_{z,n}t_n}}$$

$$t_{mo}^{TM,TE} = \frac{t_{mn}^{TM,TE} t_{no}^{TM,TE} e^{ik_{z,n}t_n}}{1 + r_{mn}^{TM,TE} r_{no}^{TM,TE} e^{2ik_{z,n}t_n}} \tag{S8}$$

It should be noted that by using (S5), the calculated heat flux from medium 1 entering mediums 3 and 4 will be obtained. In order to isolate the contribution absorbed by medium 3, the portion of energy exiting medium 3 and entering medium 4 must be subtracted. To determine this heat flux, the same procedure is applied to obtain the Poynting vector from medium 1 to medium 4. In this case, medium 4 is assumed to be semi-infinite which thus simplifies the coefficients as shown,

$$A_4^{TM,TE} = \frac{e^{ik_{z,1}t_H} e^{ik_{z,2}g} e^{ik_{z,3}t_C} t_{21}^{TM,TE} t_{23}^{TM,TE} t_{34}^{TM,TE}}{\left(1 + r_{21}^{TM,TE} r_{10}^{TM,TE} e^{2ik_{z,1}t_H}\right)\left(1 + r_{23}^{TM,TE} r_{34}^{TM,TE} e^{2ik_{z,3}t_C}\right)\left(1 - r_{20}^{TM,TE} r_{24}^{TM,TE} e^{2ik_{z,2}g}\right)} \cdot \frac{k_{z,1}}{k_{z,2}}$$

$$B_4^{TM,TE} = 0 \tag{S9}$$

$$C_4^{TM,TE} = r_{34}^{TM,TE} \cdot A_4^{TM,TE}$$

$$D_4^{TM,TE} = 0$$

To determine the net radiative heat transfer between mediums 1 and 3, the heat flux from medium 3 to medium 1 must also be found using the same procedure. By using the more general form of radiative transfer based on Rytov theory, no assumptions are made on the symmetry of the system. The model allows calculations using different emitter and PV cell materials, different film thicknesses, and different back side reflectors on the hot and cold sides.

**Supplementary Figures:**

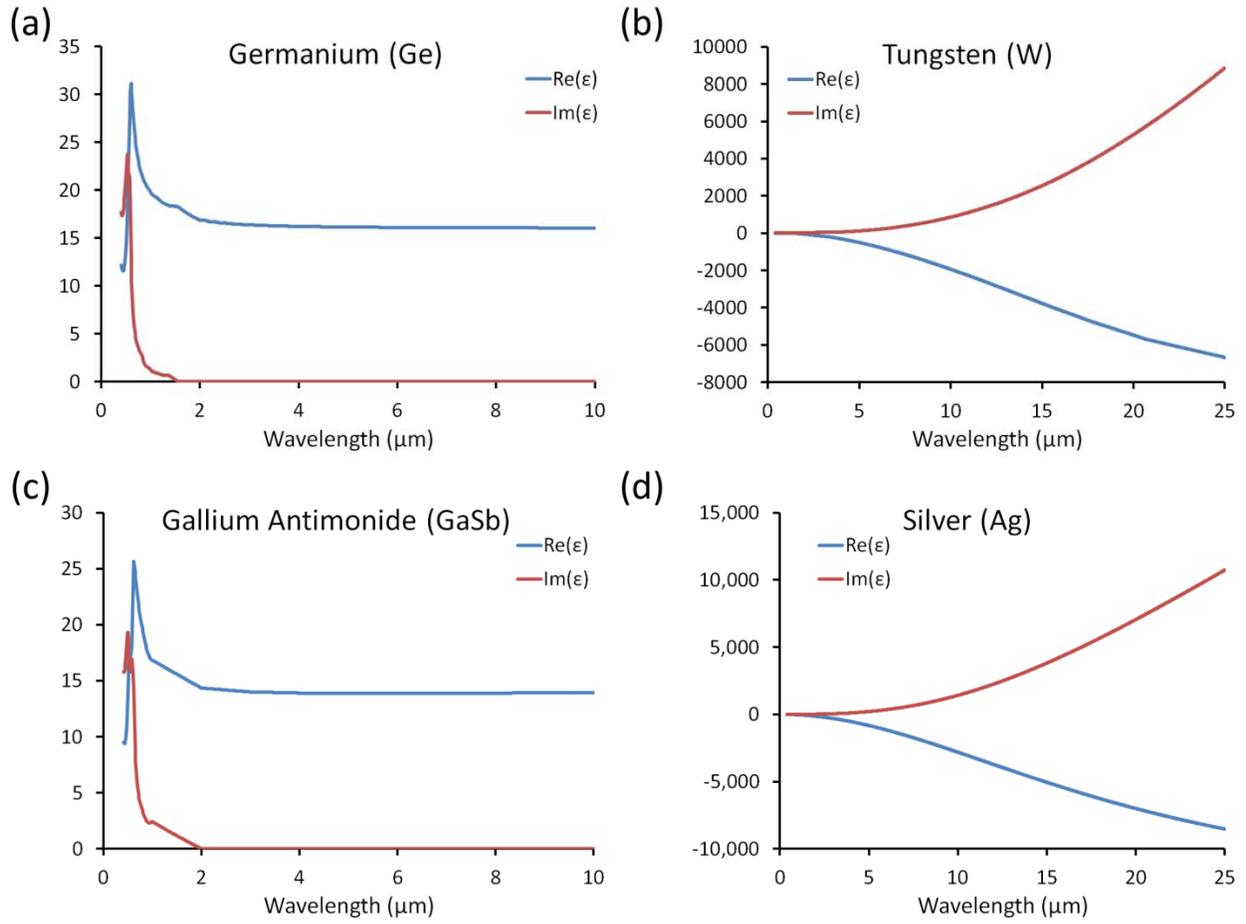

**Figure S1:** Optical constants of: (a) germanium (Ge), (b) tungsten (W), (c) gallium antimonide (GaSb), and (d) silver (Ag) obtained from the literature.[5] The optical constants for the perfect metal were taken in the long wavelength limit of silver and were assumed to be dispersionless for the wavelength range computed.

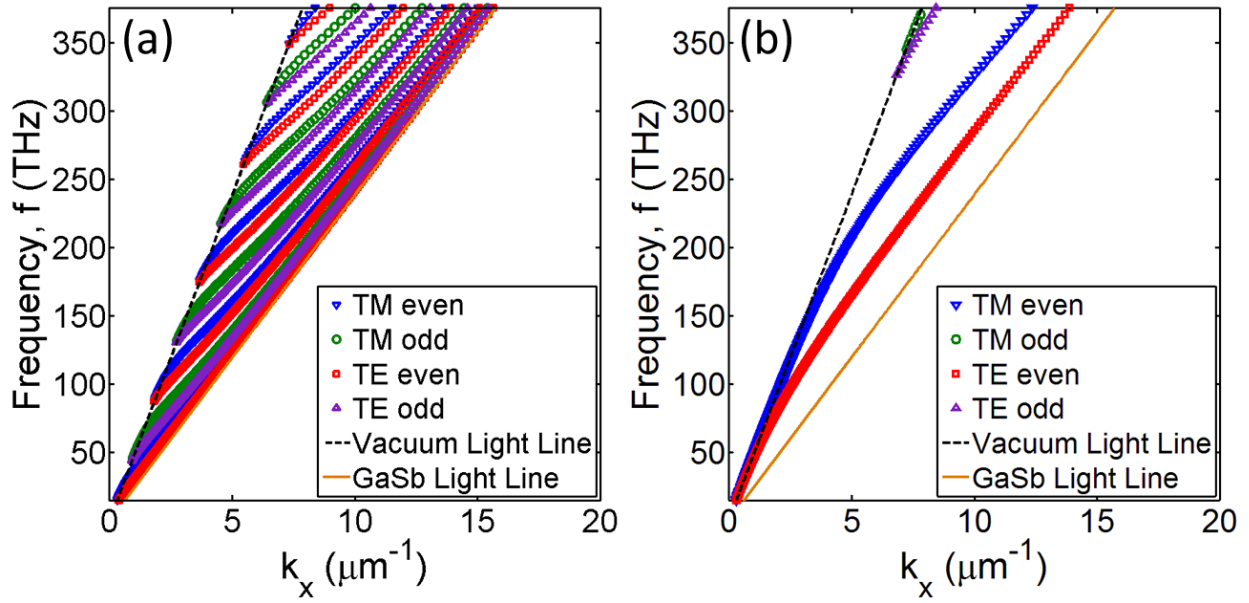

**Figure S2:** The dispersion for an isolated thin-film GaSb waveguide suspended in vacuum. (a) The dispersion for a 1 μm thick GaSb thin-film. (b) The dispersion for a 136 nm thick GaSb thin-film. As shown, variations in thickness can dramatically change the number of modes while simultaneously blue shifting the cut-off frequency of each waveguide mode. The first fundamental modes shown in (b) do not exhibit a cut-off frequency due to the symmetric nature of the waveguide. However, in the thin-film TPV system, the asymmetric nature of the waveguide system ensures a cut-off does exist.

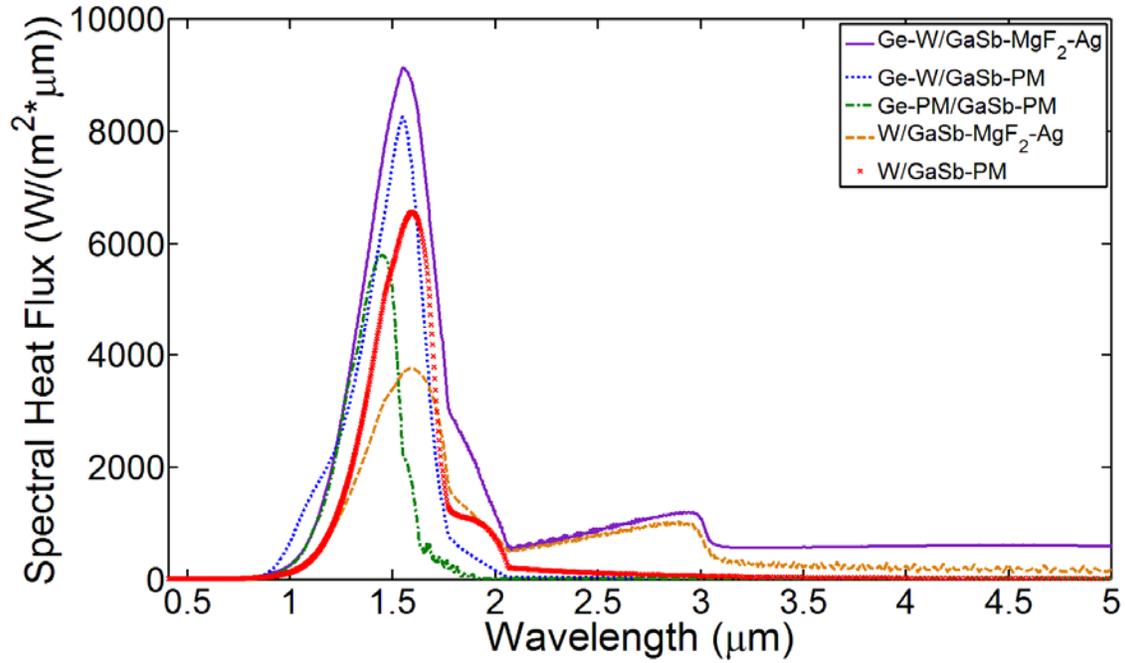

**Figure S3:** The spectral heat flux for the simulated cases presented in the main text. All cases are optimized to maximize the energy conversion efficiency. The thicknesses for all cases are: (1) Ge-W/GaSb-MgF$_2$-Ag: $t_H$ = 119 nm, $t_C$ = 100 nm, $t_S$ = 1.25 µm, (2) Ge-W/GaSb-PM: $t_H$ = 58 nm, $t_C$ = 94 nm, (3) Ge-PM/GaSb-PM: $t_H$ = 860 nm, $t_C$ = 136 nm, (4) W/GaSb-MgF$_2$-Ag: $t_C$ = 59nm, $t_S$ = 750 nm, (5) W/GaSb-PM: $t_C$ = 134 nm.

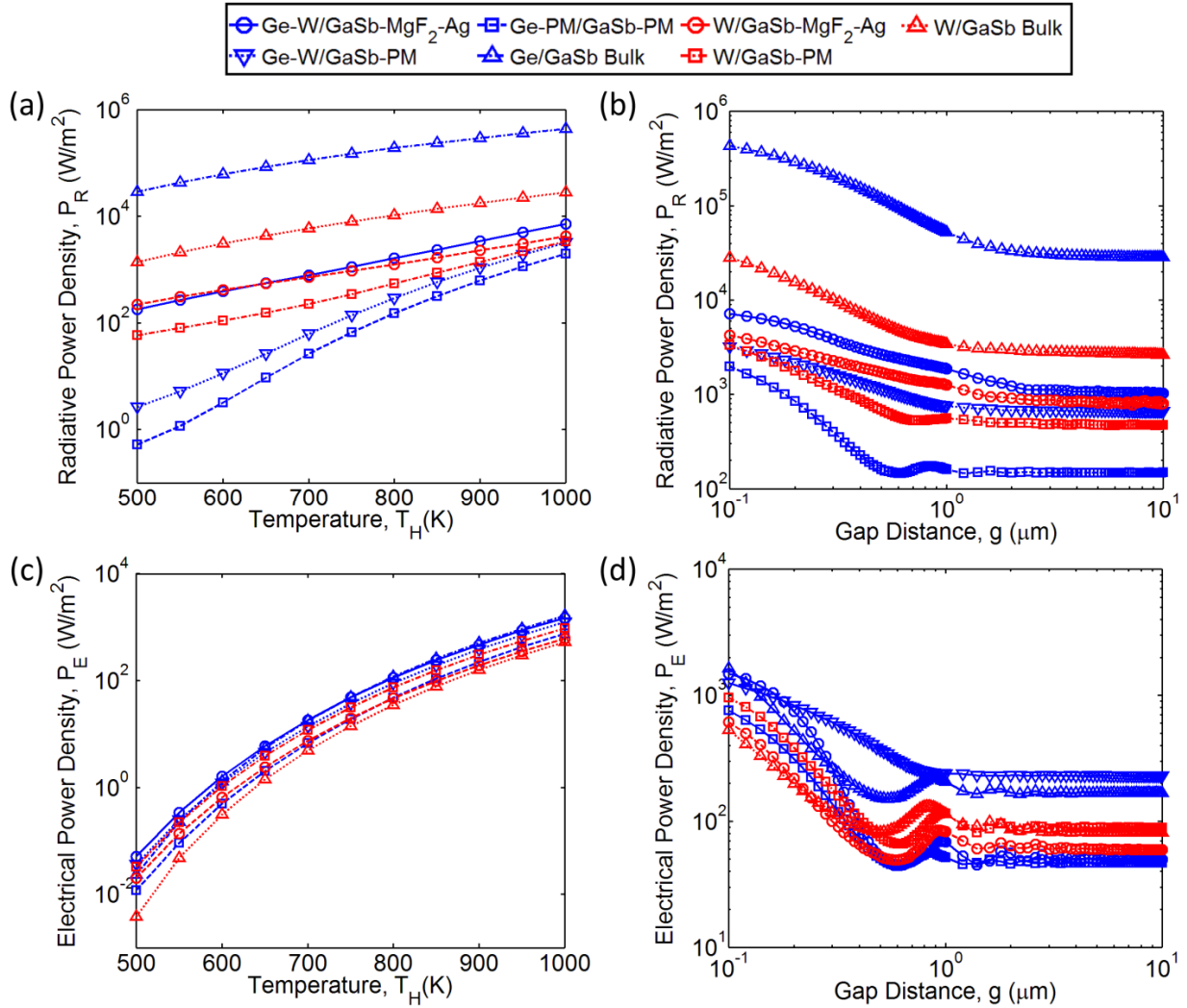

**Figure S4:** The radiative power density, $P_R$, which is defined as the net radiative transfer to the PV cell and underlying substrate for various combinations of the emitter material and back-reflector material as a function of (a) the emitter temperature $T_H$ and (b) the gap separation g between the emitter and the PV cell. The electrical power density, $P_E$, for various combinations of the emitter material and back-reflector material as a function of (c) the emitter temperature $T_H$ and (d) the gap separation g between the emitter and the PV cell.

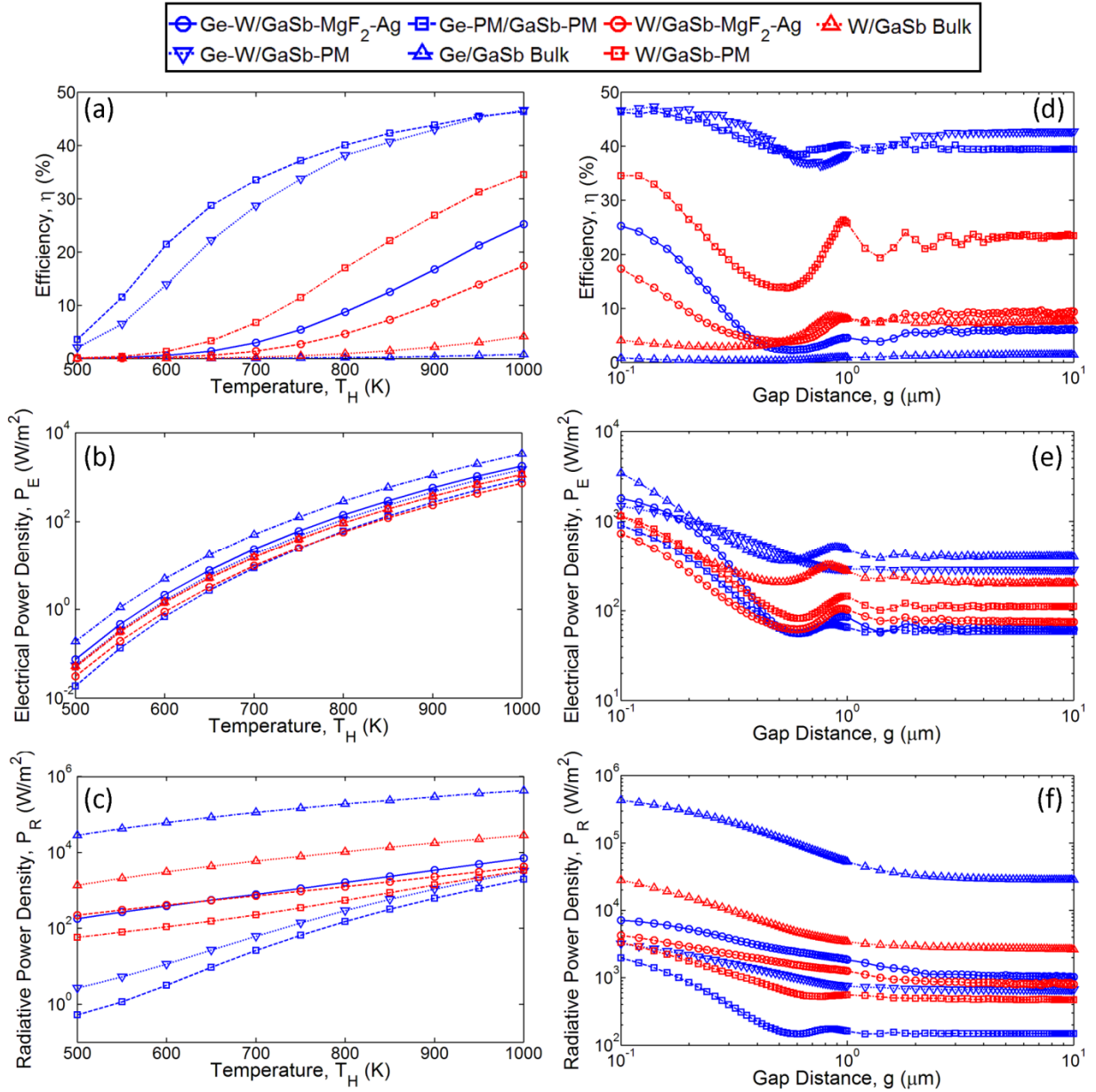

**Figure S5:** The predicted efficiency, η, the electrical power density, $P_E$, and the radiative power density, $P_R$, using the Shockley-Queisser formulation. All cases correspond to the main text using the same dimensions and parameters. Assuming a gap separation of g = 100 nm, the (a) efficiency, (b) electrical power density, and (c) radiative power density is plotted as a function of emitter temperature, $T_H$. Likewise, assuming an emitter temperature of $T_H$ = 1000 K, the (d) efficiency, (e) electrical power density, and (f) radiative power density is plotted as a function of gap separation, g.

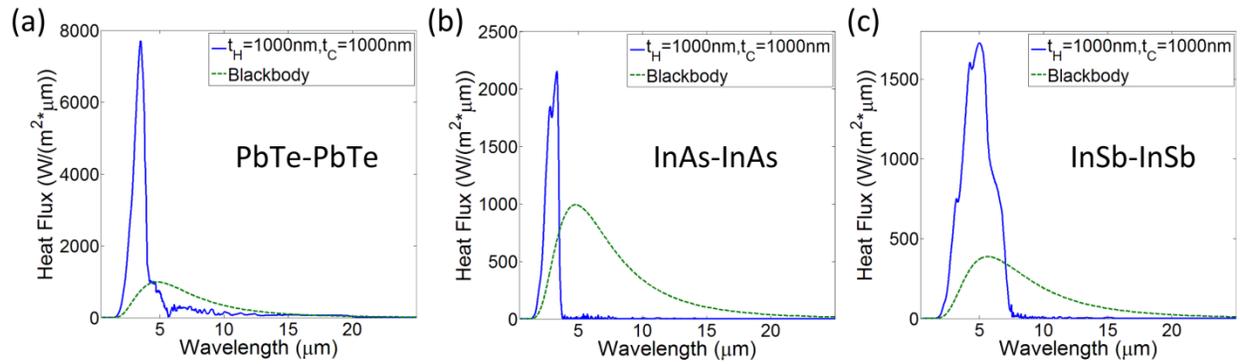

**Figure S6:** To show the generality of the 'thermal well' effect in manipulating thermal radiative transfer, several alternative material combinations were calculated, including: (a) lead telluride (PbTe) emitter and absorber assuming an emitter temperature of $T_H$ = 600 K and absorber temperature of $T_C$ = 300 K at a separation distance of 100 nm, (b) indium arsenide (InAs) emitter and absorber assuming an emitter temperature of $T_H$ = 600 K and absorber temperature of $T_C$ = 300 K at a separation distance of 100 nm, and (c) indium antimonide (InSb) emitter and absorber assuming an emitter temperature of $T_H$ = 500 K and absorber temperature of $T_C$ = 300 K at a separation distance of 100 nm. In all cases, the back-reflector is chosen to be a perfect metal. It should be noted that these materials were not intended for use in high-temperature TPV applications, but rather to assess the potential spectral selectivity of thermal emission that could be achieved using the thermal well concept with different materials.

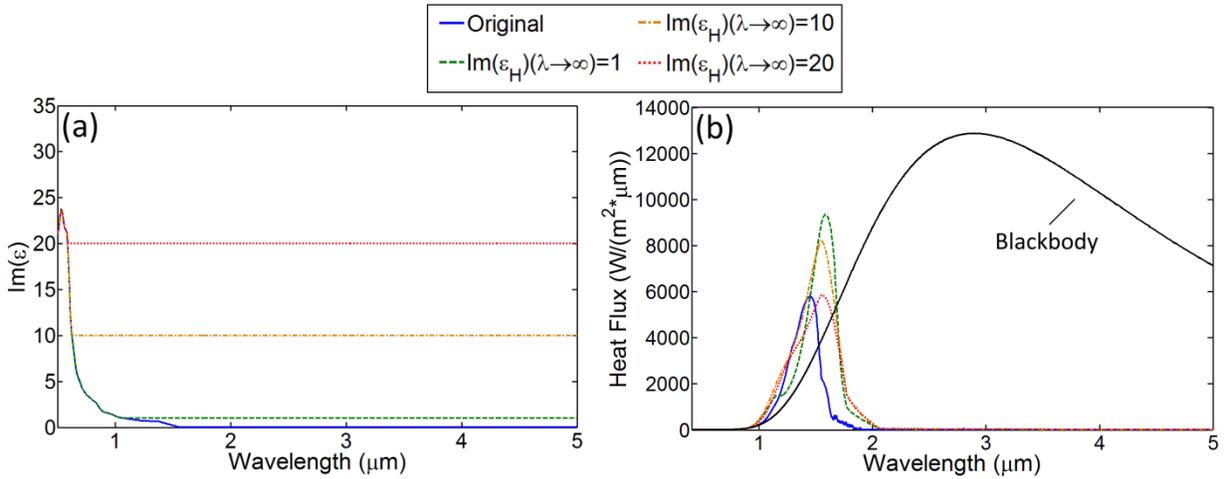

**Figure S7:** At high temperatures, the optical properties of Ge will change due to a combination of thermal expansion which decreases the electronic band-gap and a significant increase in the population of thermally excited free-carriers. These factors will both lead to emission at longer wavelengths. (a) To assess whether this will impact the spectral selectivity, the imaginary component of permittivity of germanium was artificially increased to simulate this effect. Each case was re-optimized to maximize efficiency. For $Im(\varepsilon_H \rightarrow \infty) = 1$: $t_H = 100$ nm, $t_C = 100$ nm, $Im(\varepsilon_H \rightarrow \infty) = 10$: $t_H = 90$ nm, $t_C = 100$ nm, $Im(\varepsilon_H \rightarrow \infty) = 20$: $t_H = 80$ nm, $t_C = 100$ nm (b) The spectral radiative heat flux for a Ge thin-film emitter and a GaSb PV cell supported by a perfect metal still exhibits spectrally selective radiative transfer even for high material losses in the emitter. This can be explained by the inability of the GaSb thin-film to absorb long wavelength thermal radiation emitted by Ge due to the cutoff frequency of the lowest frequency mode. Although the spectral heat flux broadens and redshifts for higher material losses, the efficiency computed for each case is 39.8% for $Im(\varepsilon_H \rightarrow \infty) = 1$, 37.7% for $Im(\varepsilon_H \rightarrow \infty) = 10$, and 36% for $Im(\varepsilon_H \rightarrow \infty) = 20$.

**References**

1. Rytov, S. M., Kravtsov, Y. A. & Tatarskii, V. I. *Principles of Statistical Radiophysics 3: Elements of Random Fields*. (Springer-Verlag Berlin Heidelberg, 1989).
2. Shen, S., Narayanaswamy, A. & Chen, G. Surface phonon polaritons mediated energy transfer between nanoscale gaps. *Nano Lett.* **9,** 2909–2913 (2009).
3. Francoeur, M., Mengüç, M. P. & Vaillon, R. Spectral tuning of near-field radiative heat flux between two thin silicon carbide films. *J. Phys. D. Appl. Phys.* **43,** 075501 (2010).
4. Francoeur, M., Pinar Mengüç, M. & Vaillon, R. Solution of near-field thermal radiation in one-dimensional layered media using dyadic Green's functions and the scattering matrix method. *J. Quant. Spectrosc. Radiat. Transf.* **110,** 2002–2018 (2009).
5. Palik, E. D. *Handbook of optical constants of solids*. (Academic Press, 1997).